\documentclass[12pt]{iopart}
\begin{document}
%
\title[The interplay of supersymmetry and ${\cal  PT}$ symmetry]
{The interplay of supersymmetry and ${\cal  PT}$ symmetry
in quantum mechanics: a case study for the Scarf II potential}

\author{G. L\'evai\dag\footnote[3]{E-mail: levai@atomki.hu} and 
        M. Znojil\ddag\footnote[4]{E-mail: znojil@ujf.cas.cz\ ,
	http://gemma.ujf.cas.cz/~znojil}}

\address{\dag\ Institute of Nuclear Research of the Hungarian
         Academy of Sciences, \\
         PO Box 51, H--4001 Debrecen, Hungary}
\address{\ddag\ Nuclear Physics Institute of Academy of Sciences of
         the Czech Republic, \\ 
	 250 68 \v Re\v z, Czech Republic}

\begin{abstract}
Motivated by the duality of normalizable states and the presence of 
the quasi-parity quantum number $q=\pm 1$ in ${\cal  PT}$ symmetric 
(non-Hermitian) quantum mechanical potential models, the relation of 
${\cal PT}$ symmetry and supersymmetry (SUSY) is studied.  As an 
illustrative example the ${\cal  PT}$ invariant version of the Scarf 
II potential is presented, and it is shown that the ``bosonic" 
Hamiltonian has two {\it different} ``fermionic" SUSY partner 
Hamiltonians (potentials) generated from the ground-state solutions 
with $q=1$ and $q=-1$. It is shown that the ``fermionic" potentials 
cease to be ${\cal  PT}$ invariant when the ${\cal PT}$ symmetry of 
the ``bosonic" potential is spontaneously broken. A modified 
${\cal  PT}$ symmetry inspired SUSY construction is also discussed, 
in which the SUSY charge operators contain the antilinear operator 
${\cal  T}$. It is shown that in this scheme the ``fermionic" 
Hamitonians are just the complex conjugate of the original ``fermionic" 
Hamiltonians, and thus possess the same energy eigenvalues. 
\end{abstract}


\maketitle

\section{Introduction}

Symmetries and invariance properties are among the most
characteristic features of any physical system. They usually give
a deeper insight into the physical nature of the problem, but also
help their mathematical formulation. Symmetries typically lead to
characteristic patterns in the energy spectrum of the system.
These features are shared by the ``classic'' potential problems of
non-relativistic quantum mechanics. Technically these are
relatively simple systems, and accordingly they include a number
of exactly solvable examples, nevertheless, they represent the
showcase of a wide variety of symmetry and invariance concepts.
The most widely known symmetries of quantum mechanical potentials
are based on group theory (in particular, Lie algebras),
supersymmetry and ${\cal  PT}$ symmetry.

Group theoretical approaches to quantum mechanical problems and
potentials in particular, are practically as old as quantum
mechanics itself. The elements of the (symmetry, 
spectrum generating, dynamical \cite{algebras} and 
potential \cite{potalg}) algebras typically connect different
eigenstates of the same Hamiltonian or some interrelated
Hamiltonians, while the states themselves belong to the
irreducible representations of the corresponding group.

A less immediate application of the concept of symmetry appears in
supersymmetric quantum mechanics (SUSYQM)
\cite{susy}, where the supersymmetry relates two Hamiltonians
which typically have identical spectra except, possibly, the 
ground state of one of the Hamiltonians which is missing from the 
spectrum of the other one. For traditional reasons these two 
Hamiltonians are called the ``bosonic'' and the ``fermionic'' 
Hamiltonians and are denoted with the ``--'' and ``+'' indices. In 
SUSYQM they are constructed from two linear first-order 
differential equations as $H_- = A^\dagger A$ and 
$H_+ = A\,A^\dagger$. 
SUSYQM is essentially a reformulation of the factorization
technique which is an old method of generating isospectral
potentials \cite{ih51}. 

The most recent symmetry concept is the so called ${\cal  PT}$
symmetry of one-dimensional quantum mechanical potentials. It has
several relevant and interesting implications regarding the energy
spectrum. In ${\cal PT}$ symmetric quantum mechanics of Bender and
Boettcher \cite{bb98} the potentials are invariant under the
simultaneous action of the space and time reflection operations
${\cal  P}$ and ${\cal  T}$, and have the property
$[V(-x)]^*=V(x)$. A peculiar feature of these models is that
although they are not Hermitian, they may possess real
bound-state energy spectrum. Alternatively, the ${\cal  PT}$
symmetric potentials may support eigenvalues arranged into 
complex conjugate pairs \cite{AM}, but then the energy
eigenfunctions cease to be eigenfunctions of the ${\cal PT}$
operator, and the emergence of a complex energy can be interpreted
as a manifestation of the spontaneous breakdown of ${\cal  PT}$
symmetry.

In the above symmetry-based scenario occurring in numerous
applications of quantum mechanics an interplay may be noticed
between different symmetry concepts. For example, the 
practical identity of the supersymmetric shift operators $A$ and 
$A^{\dag}$ with the ladder operators of some potential algebras 
has been established \cite{jpa94} in the case of potential 
families corresponding to type A and B factorizations 
\cite{ih51}. Similarly, in spite of the comparative novelty of the
${\cal PT}$ symmetric quantum mechanics, some standard
Lie-algebraic methods found already their inspiring applications
within its non-standard framework. Some solvable ${\cal  PT}$
symmetric potentials have been associated with the sl(2,$C$) 
\cite{bbcq00,bbcq02}, su(1,1)$\simeq$so(2,1) \cite{lgfcav01} 
and so(2,2) \cite{lgfcav02} potential algebras.

The relationship between the supersymmetric and ${\cal  PT}$
symmetric considerations may still be felt as a certain ``missing
link". This motivated our forthcoming analysis. Firstly, on the
background of \cite{qtsusy} we imagined that there exist
essential differences between any Hermitian and non-Hermitian
versions of the supersymmetric formalism. We believe that it
deserves a deeper study, first of all, via particular examples.
Secondly, we were always aware that more attention has to be paid
to one of the most specific features of the ${\cal  PT}$ symmetric
Hamiltonians, namely, to the existence of the so called
quasi-parity quantum number $q$ which, roughly speaking, reflects
the emergence of new normalizable states during the transition
from the Hermitian to non-Hermitian $H$. 
This extension of the basis states has already been investigated in 
an algebraic formalism \cite{lgfcav02}. Last but not least, we
were encouraged by the increasing number of the available ${\cal 
PT}$ symmetric examples where the reality of spectra was explained
using techniques of SUSYQM~\cite{cjt98,aicd99,bbrr00,tateo,bmq02}.

In our paper we shall pay more attention to the recent observation
\cite{mzjpa02} that the quasi-parity may play a key role in the
latter context. We are going to emphasize that in a way which
extends the scope of the latter reference, the formal changes of
the SUSYQM rules in the non-Hermitian case are {\em not} an
artifact of the presence of the singularities in the complex plane
of $x$. For the sake of clarity of our argument we shall pick up
first one characteristic potential (often called Scarf II) and
summarize its known properties in section 2. Section 3 is then 
devoted to its deeper analysis. We shall see that the
supersymmetrization of this potential in its fully regular 
${\cal PT}$ symmetric version exhibits some properties which make 
it very different from its stardard treatment within the Hermitian 
SUSYQM. The respective discussion and summary of our findings are 
finally collected in sections 4 and~5.

\section{The Scarf II potential and its ${\cal PT}$ symmetric 
version}

In a way which reflects the innovative character of the ${\cal 
PT}$ symmetric models, many of their studies focused on a
particular potential. 
The first examples of these potentials have
even been found using perturbation techniques \cite{pert}. 
Further ones have been identified using semiclassical
approximations \cite{bb98,semi} and numerical algorithms
\cite{num}. A number of the exactly solvable ${\cal PT}$ symmetric
potentials have been revealed as the analogues of their Hermitian,
real special cases \cite{cjt98,bbrr00,mz,mzb,lgmz00}. In such a
setting we shall pick up the Scarf II potential
\begin{equation}
\fl
V(x)=-\frac{1}{\cosh^2 x}
\left[\left(\frac{\alpha+\beta}{2}\right)^2
+\left(\frac{\alpha-\beta}{2}\right)^2  -\frac{1}{4}\right]
+\frac{2{\rm i}\sinh x}{\cosh^2 x}
\left(\frac{\beta+\alpha}{2}\right)
\left(\frac{\beta-\alpha}{2}\right) 
\label{vscarf}
\end{equation}
as a typical illustration of the generic relations between the
concept of ${\cal PT}$ symmetry and certain SUSYQM constructions.
The Scarf II potential seems to be an ideal example for such a
purpose since

\begin{itemize}
\item
it is one of the shape-invariant potentials \cite{si} which, in
the SUSYQM context, belongs to type A factorization \cite{ih51};
\item
its general functional form contains its Hermitian version 
(for $\alpha=\beta^*$ \cite{dab88,lg89}) as well as its 
${\cal  PT}$ symmetric one (for $\alpha$ and $\beta$ both real 
or imaginary \cite{lgmz00,lgmz01}); 
\item
in contrast with many ${\cal  PT}$ symmetric potentials generated
from the singular Hermitian potentials, its analysis
\cite{bbrr00,mzb} does not require any artificial regularization;
\item
the spontaneous breakdown of its ${\cal  PT}$ symmetry
\cite{bbcq02,lgmz01,ahmed01a,bbcqmz01,pla02} occurs simply due to
a change of one of its real parameters to an imaginary value.
\end{itemize}

Before addressing the details we note that our notation can easily 
be transformed to that used in the other works. Thus, we might put
$V_1=[(\alpha+\beta)^2 +(\alpha-\beta)^2 -1]/4$ and
$V_2=(\alpha+\beta)(\alpha-\beta)/2$ in \cite{ahmed01a,bbcq02},
$A=-(\alpha+\beta+1)/2$ and $B=(\alpha-\beta)/2$ in 
\cite{bbcqmz01} and $s=-(\alpha+\beta+1)/2$ and $\lambda={\rm
i}(\alpha-\beta)/2$ in \cite{dab88,lg89}.

\subsection{The Hermitian Scarf II potential}

The conventional Hermitian version of the Scarf II potential 
\cite{dab88,lg89} is obtained when the second term in (\ref{vscarf}) 
is made real by the $\alpha=\beta^*=-s-\frac{1}{2}-{\rm i}\lambda$ 
parametrization, for example. One then finds the bound-state 
(normalizable) solutions at the energies
\begin{equation}
E_n=-\left(n+\frac{\alpha+\beta+1}{2}\right)^2\ ,
 \label{oescarf}
\end{equation}
with the corresponding wavefunctions expressed in terms of 
Jacobi polynomials as 
\begin{equation}
\psi_n(x) =C_n (1-{\rm i}\sinh x)^{\frac{\alpha}{2}+\frac{1}{4}}
 (1+{\rm i}\sinh x)^{\frac{\beta}{2}+\frac{1}{4}}
P_n^{(\alpha,\beta)}({\rm i}\sinh x)\ . \label{ofscarf}
\end{equation}
We note that although the Scarf II (or Gendenshtein) potential has 
been known for some time, the normalization constant $C_n$ were 
determined only recently in \cite{pla02}.  
The condition of the normalizability of the (\ref{ofscarf}) 
functions limits the range of the admissible quantum numbers via 
\begin{equation}
n<-[{\rm Re}(\alpha+\beta)+1]/2\ . \label{oncond}
\end{equation}

We may note that the $\alpha\leftrightarrow\beta$ transformation
changes the sign of the odd component of (\ref{vscarf}) and leaves
the even one invariant. This mimics the spatial reflection
operation ${\cal  P}$ and has no effect on the energy spectrum. A
simple calculation based on the properties of Jacobi polynomials
reveals that the interchange $\alpha\leftrightarrow\beta$ acts as
a spatial reflection on the wavefunctions (\ref{ofscarf}), up to
an unimportant sign change $(-1)^n$ in their norm~\cite{pla02}.

We note that equations (\ref{vscarf}), (\ref{oescarf}), 
(\ref{ofscarf}) and (\ref{oncond}) apply to the general complex 
version of the Scarf II potential too, although in this case further 
refining of the formalism might become necessary, as we shall see in 
the next subsection.

\subsection{The ${\cal  PT}$ symmetric Scarf II potential and the 
quasi-parity}

In the light of the review \cite{lgmz00} the ${\cal  PT}$
symmetric version of the Scarf II potential is obtained if
$\alpha^*=\pm\alpha$ and $\beta^*=\pm\beta$ holds, i.e. if
$\alpha$ and $\beta$ are both either real or imaginary. The energy
eigenvalues are all real, and the ${\cal PT}$ symmetry is
unbroken, if both $\alpha$ and $\beta$ are real. When one of the
two parameters is real and the other one is imaginary, then the
energy eigenvalues appear in complex conjugate pairs, and this
case corresponds to the spontaneous breakdown of ${\cal  PT}$
symmetry \cite{lgmz01,pla02}. If both parameters are imaginary, 
then there are no normalizable states due to the 
constraint~(\ref{oncond}).

Based on the practical equivalence of $\alpha$ and $\beta$, we can
assume without any loss of generality that $\beta$ is real and
$\alpha$ is either real or imaginary, depending on whether we
study unbroken or spontaneously broken ${\cal  PT}$ symmetry. A
remarkable feature of ${\cal  PT}$ symmetric potentials is that
the range of their normalizable states is broader than that of
their Hermitian counterparts. In case of the Scarf II potential
this is reflected by the fact that both $\alpha$ and
$-\alpha$ can appear in Eqs. (\ref{ofscarf}), (\ref{oescarf}) and
(\ref{oncond}), since the potential (\ref{vscarf}) is invariant
under the $\alpha\rightarrow -\alpha$ transformation.

The dual admissible sign of $\alpha$ may be called a quasi-parity
quantum number \cite{mz,bbcqmz01,bbsmcq01} $q=\pm 1$. This makes 
the Scarf II potential similar to other ${\cal  PT}$ symmetric 
potentials, which also have a second set of bound-state solutions 
compared to their Hermitian versions. However, the mechanism of the
appearance of the second set is different from the scenario
typical for the singular potentials where the singularity is
cancelled by the ${\cal  PT}$ symmetric regularization procedure.
Now the new states ``evolve" from states that already existed as
resonances in the Hermitian limit \cite{lgfcav01}, so that their 
emergence is not directly related to the less strict boundary 
conditions.

In what follows we shall modify
our notation slightly and replace $\alpha$ with $q\alpha$, where
$q=\pm 1$ is the quasi-parity. This implies a redefinition of the
formulae used previously as

\begin{equation}
E^{(q)}_n=-\left(n+\frac{q\alpha+\beta+1}{2}\right)^2\ ,
\label{eqscarf}
 \label{escarf}
\end{equation}

\begin{equation}
\psi^{(q\alpha,\beta)}_n(x) =C^{(q)}_n (1-{\rm i}\sinh
x)^{\frac{q\alpha}{2}+\frac{1}{4}}
 (1+{\rm i}\sinh x)^{\frac{\beta}{2}+\frac{1}{4}}
P_n^{(q\alpha,\beta)}({\rm i}\sinh x)\ . \label{fqscarf}
\label{fscarf}
\end{equation}
The number of bound states contained in the two sets depends on
$q$ because the condition for normalizability is also modified
accordingly,
\begin{equation}
n<-[{\rm Re}(q\alpha+\beta)+1]/2\ . \label{ncond} \label{qncond}
\end{equation}
In particular, when both $\alpha$ and $\beta$ are real, i.e. the
${\cal  PT}$ symmetry is unbroken, the bounds (\ref{qncond})
differ in general, however, in the case of spontaneously broken
${\cal PT}$ symmetry, when $\alpha$ is imaginary, the two sets
contain equal number of bound states, as expected from the fact
that the two sets are formed by complex conjugate energy
eigenvalues.

In \cite{pla02} the normalization constants in (\ref{fscarf}) 
have been determined using the modified inner product 
$\langle \psi\vert {\cal  P}\vert \psi \rangle$ of 
\cite{mz0103054}. The (\ref{fscarf}) functions were found to be 
orthogonal to each other using this inner product, while (as 
expected for a non-Hermitian problem), this was not the case when 
the standard Hermitian inner product was used.

\section{Supersymmetrization of the ${\cal  PT}$ invariant
Scarf II potential}

The duality of normalizable solutions in the ${\cal  PT}$
symmetric setting implies a duality in the superpotentials too,
and this is a remarkable new feature of the supersymmetrization of
${\cal  PT}$ symmetric potentials. This means that in the
realization of the standard $N=2$ SUSYQM algebra \cite{susy}
\begin{equation}
\fl
\{Q,Q^{\dag}\}=0 \hskip 1cm \{Q,Q\}=\{Q^{\dag},Q^{\dag}\}=0 \hskip
1cm [{\cal  H},Q]=[{\cal  H},Q^{\dag}]=0 \label{ssalg}
\end{equation}
the supersymmetric charge operators
\begin{equation}
Q=\left(
\begin{array}{cc} 0 & 0 \\
A^{(q)} & 0 \\
\end{array}\right)
\hskip 1.5cm
Q^{\dag}=\left(
\begin{array}{cc} 0 & A^{\dag (q)}\\
0 & 0 \\
\end{array}\right)
\label{scharge}
\end{equation}
and the supersymmetric Hamiltonian
\begin{equation}
{\cal  H}=\left(
\begin{array}{cc} H_-^{(q)} & 0 \\
0 & H_+^{(q)} \\
\end{array}\right)
\equiv
\left(
\begin{array}{cc}  A^{\dag (q)}A^{(q)} & 0 \\
0 &  A^{(q)}A^{\dag (q)}  \\
\end{array}\right)
\label{sham}
\end{equation}
are constructed using the SUSYQM shift operators
\begin{equation}
A^{(q)}=\frac{\rm d}{{\rm d}x}+W^{(q)}(x) \hskip 1.5cm
A^{\dag(q)}=-\frac{\rm d}{{\rm d}x}+W^{(q)}(x)
\label{sslad}
\end{equation}
which now depend explicitly on the quasi-parity quantum number
$q=\pm 1$:
\begin{eqnarray}
W^{(q)}(x)&=&-\frac{\rm d}{{\rm d}x}\ln 
\psi^{(q\alpha,\beta)}_{0,-}(x)
\nonumber\\
&=&-\frac{1}{2}(q\alpha+\beta+1)\tanh x
-\frac{\rm i}{2}(\beta-q\alpha)\frac{1}{\cosh x}\ ,
\label{wscarf}
\end{eqnarray}
where $\psi^{(q\alpha,\beta)}_{n,-}(x)=
\psi^{(q\alpha,\beta)}_n(x)$, so $\psi^{(q\alpha,\beta)}_{0,-}(x)$
is the ground-state wavefunction of the ``bosonic" Hamiltonian 
$H_-^{(q)}$.

The``bosonic" potential may be constructed by using the standard
SUSYQM recipe,
\begin{eqnarray}
U^{(q)}_-(x)&=&[W^{(q)}(x)]^2 -\frac{{\rm d}W^{(q)}}{{\rm d}x}
\label{vm}
\\
&=&-\frac{1}{\cosh^2 x}
\left[\left(\frac{q\alpha+\beta}{2}\right)^2
+\left(\frac{q\alpha-\beta}{2}\right)^2  -\frac{1}{4}\right]
\nonumber\\ && +\frac{2{\rm i}\sinh x}{\cosh^2 x}
\left(\frac{\beta+q\alpha}{2}\right)
\left(\frac{\beta-q\alpha}{2}\right)
+\left(\frac{q\alpha+\beta+1}{2}\right)^2\ . \label{vmscarf}
\end{eqnarray}
This expression gives the same potential (\ref{vscarf}) as before,
except for a constant term, which is simply an energy shift
securing zero ground-state energy. In order to get back the
original potential (\ref{vscarf}) and energy eigenvalues
(\ref{escarf}) we have to shift the energy scale, {\em subtracting the
$q$-dependent constant term} from the potential (and the energy
expression) obtained in the SUSYQM procedure:
\begin{equation}
V^{(q\alpha,\beta)}_-(x)\equiv
U^{(q)}_-(x) -\left(\frac{q\alpha+\beta+1}{2}\right)^2
=V(x)\ ,
\label{vvmscarf}
\end{equation}
\begin{equation}
E^{(q\alpha,\beta)}_{n,-}=E^{(q)}_n\ .
\label{emscarf}
\end{equation}
Note that $V^{(q\alpha,\beta)}_-(x)=V(x)$ does not depend on $q$,
as we have established previously. In what follows, therefore, 
we shall introduce the notation ${\bf H}_-$ for the ``bosonic'' 
Hamiltonian in which the relative energy shift has been applied.  
In addition, the SUSYQM
partners of the $U^{(q)}_-(x)$ potentials,
\begin{equation}
U^{(q)}_+(x)=[W^{(q)}(x)]^2 +\frac{{\rm d} W^{(q)}}{{\rm d}x}
\label{vp}
\end{equation}
contain the same energy constant as the $U^{(q)}_-(x)$ potentials
(\ref{vmscarf}). Subtracting these constants we get {\em two separate}
potentials, which depend on $q$:
\begin{eqnarray}
V^{(q\alpha,\beta)}_+(x)&\equiv& U^{(q)}_+(x)
-\left(\frac{q\alpha+\beta+1}{2}\right)^2 \nonumber\\
&=&-\frac{1}{\cosh^2 x}
\left[\left(\frac{q\alpha+\beta+2}{2}\right)^2
+\left(\frac{q\alpha-\beta}{2}\right)^2  -\frac{1}{4}\right]
\nonumber\\ && +\frac{2{\rm i}\sinh x}{\cosh^2 x}
\left(\frac{\beta+q\alpha+2}{2}\right)
\left(\frac{\beta-q\alpha}{2}\right) \nonumber\\
&&=V^{(q\alpha+1,\beta+1)}_-(x)\ . \label{vpq}
\end{eqnarray}
The main meaning of this formula is that {\em it assigns two
different} supersymmetric partners to our original Scarf II
potential and the corresponding Hamiltonian ${\bf H}_-$. We denote 
them with ${\bf H}^{(q)}_+$. 
Both potentials depend on the quasi-parity $q$ and both of
them have a shape of the Scarf II potential, with the parameters
$q\alpha$ and $\beta$ shifted by one unit. This is one of our main
observations which extends the concept of shape-invariance
\cite{si} to ${\cal PT}$ symmetric potentials.

The actual effect of the SUSYQM shift operators on the ``bosonic"
and ``fermionic" eigenfunctions can be proven by straightforward
but tedious calculations:
\begin{equation}
\fl
A^{(q)}\psi^{(q\alpha,\beta)}_{n,-}(x)
=A^{(q)}\psi^{(q\alpha,\beta)}_n(x)
\rightarrow
\psi^{(q\alpha+1,\beta+1)}_{n-1}(x)
=\psi^{(q\alpha,\beta)}_{n-1,+}(x)\ ,
\label{apm}
\end{equation}
\begin{equation}
\fl
A^{\dag (q)}\psi^{(q\alpha,\beta)}_{n-1,+}(x)
=A^{\dag (q)}\psi^{(q\alpha+1,\beta+1)}_{n-1}(x)
\rightarrow
\psi^{(q\alpha,\beta)}_n(x)
=\psi^{(q\alpha,\beta)}_{n,-}(x)\ .
\label{adpp}
\end{equation}
According to (\ref{ncond}) the two partner potentials have one
less bound (normalizable) state than the original Scarf II
potential (\ref{vscarf}), and a comparison with (\ref{eqscarf})
leads to the standard SUSYQM result for the energy eigenvalues of
the ``fermionic" sector:
\begin{equation}
E^{(q\alpha,\beta)}_{n,+}=
-\left(\frac{q\alpha+\beta+3}{2}+n\right)^2
=E^{(q)}_{n+1}\equiv
E^{(q\alpha,\beta)}_{n+1,-}\ .
\label{epscarf}
\end{equation}

Making use of the richer combination of SUSY shift operators and 
wavefunctions, we can analyse the effect of the $A^{(q)}$ 
operator on the ``bosonic'' eigenfunctions with opposite 
quasi-parity $-q$: 
\begin{equation}
\fl
A^{(q)} \psi^{(-q\alpha,\beta)}_{n,-}(x)
=A^{(q)} \psi^{(-q\alpha,\beta)}_n(x)
\rightarrow 
\psi^{(-q\alpha-1,\beta+1)}_n(x)
=\psi^{(-q\alpha,\beta)}_{n,+}(x)\ .
\label{xapm}
\end{equation}
This result indicates that the ``fermionic'' potential 
$V^{(-q\alpha,\beta)}_+(x)=V^{(-q(\alpha+1),\beta+1)}_-(x)=
V^{(q(\alpha+1),\beta+1)}_-(x)$ has the same number of states with 
$-q$ as the original bosonic potential, and this is confirmed by 
a comparison between their spectra:
\begin{equation}
E^{(-q\alpha,\beta)}_{n,+}=
-\left(\frac{-q\alpha+\beta+1}{2}+n\right)^2
=E^{(-q)}_{n}\equiv
E^{(-(q\alpha+1),\beta+1)}_{n,-}\ .
\label{xepscarf}
\end{equation}
The inverse operation of (\ref{xapm}) is 
\begin{equation}
\fl
A^{\dag (q)} \psi^{(-q\alpha,\beta)}_{n,+}(x)
=A^{\dag (q)} \psi^{(-q(\alpha+1),\beta+1)}_n(x)
\rightarrow 
\psi^{(-q\alpha,\beta)}_n(x)
=\psi^{(-q\alpha,\beta)}_{n,-}(x)\ . 
\label{xadpp}
\end{equation}
The situation is schematically illustrated in figure 1. 

The results in (\ref{xapm}) and (\ref{xadpp}) are analogous to 
some relations found in a SUSYQM inspired study of the ${\cal PT}$ 
symmetric spiked harmonic oscillator \cite{mzjpa02}, and they are 
practically equivalent with the relations describing the effect of 
the two sets of so(2,1) generators on the two sets of solutions 
of the ${\cal PT}$ symmetric Scarf II potential in the algebraic 
analysis \cite{lgfcav02}.

It has to be added though, that the new degeneracy patterns and 
partnerships of potentials are not specific to the Scarf II potential, 
rather they are valid for any system where supersymmetry and 
${\cal PT}$ symmetry appear simultaneously. To show this, we assume  
that similarly to the situation for the Scarf II potential, the 
``bosonic'' Hamiltonian ${\bf H}_-$ can be made independent of 
the quasi-parity quantum number by applying an appropriate relative 
energy shift of the $q=+1$ and $q=-1$ sectors, i.e. 
${\bf H}_-=A^{\dag (q)}A^{(q)}-\varepsilon^{(q)} 
= A^{\dag (-q)}A^{(-q)}-\varepsilon^{(-q)}$. 
With this assumption the isospectrality of the two ``fermionic'' 
Hamiltonians follows from the eigenvalue equations 
\begin{equation}
{\bf H}_- \psi^{(q)}_{n,-} = 
[A^{\dag (\pm q)}A^{(\pm q)}-\varepsilon^{(\pm q)}]\psi^{(q)}_{n,-}
= E^{(q)}_{n,-} \psi^{(q)}_{n,-}
\label{hmeig}
\end{equation}
\begin{equation}
{\bf H}^{(\pm q)}_+ \psi^{(q)}_{n,+} = 
[A^{(\pm q)}A^{\dag (\pm q)}-\varepsilon^{(\pm q)}] \psi^{(q)}_{n,+}
= E^{(q)}_{n,+} \psi^{(q)}_{n,+}\ .
\label{hpeig}
\end{equation}
From (\ref{hmeig}) it follows that the energy shift is related to 
the ``bosonic'' ground-state energy as 
$\varepsilon^{(\pm q)}=-E^{(\pm q)}_{0,-}$, as indeed was the case 
for the Scarf II potential. (See (\ref{escarf}), (\ref{vvmscarf}) 
and (\ref{emscarf}).)
With the eigenvalue equations above, 
\begin{eqnarray}
A^{(q)} {\bf H}_- \psi^{(q)}_{n,-} &=& 
A^{(q)} [A^{\dag (q)}A^{(q)}-\varepsilon^{(q)}]\psi^{(q)}_{n,-} 
\nonumber\\
&=& {\bf H}_+^{(q)} A^{(q)} \psi^{(q)}_{n,-} 
\nonumber\\
&=& E^{(q)}_{n,-} A^{(q)} \psi^{(q)}_{n,-}\ , 
\label{oldeig}
\end{eqnarray}
and 
\begin{eqnarray}
A^{(-q)} {\bf H}_- \psi^{(q)}_{n,-} &=& 
A^{(-q)} [A^{\dag (-q)}A^{(-q)}-\varepsilon^{(-q)}]\psi^{(q)}_{n,-} 
\nonumber\\
&=& {\bf H}_+^{(-q)} A^{(-q)} \psi^{(q)}_{n,-} 
\nonumber\\
&=& E^{(q)}_{n,-} A^{(-q)} \psi^{(q)}_{n,-}\ , 
\label{neweig}
\end{eqnarray}
so we find that the $A^{(\pm q)} \psi^{(q)}_{n,-}$ functions are 
eigenfunctions of the ${\bf H}_+^{(\pm q)}$ ``fermionic'' Hamiltonians, 
and the corresponding energy eigenvalues are the same as those of 
the $q$-independent ``bosonic'' Hamiltonian. There is a difference 
between (\ref{oldeig}) and (\ref{neweig}) that in the former case 
$A^{(q)} \psi^{(q)}_{n,-}=0$ by construction, so the partner of 
the ground-state ``bosonic'' level is missing from the ``fermionic'' 
Hamiltonian ${\bf H}_+^{(q)}$, while the situation is different for 
(\ref{neweig}), so there the number of levels is the same in the 
``bosonic'' and ``fermionic'' Hamiltonians, just as we have seen 
for the example of the Scarf II potential. 
These considerations prove that the combination of the two symmetries 
leads to a richer spectral pattern than either of them separately, not 
only for the Scarf II potential, but also for any $q$-independent 
potentials.

\section{Discussion}

After one moves from the analytic to algebraic context the
doubling of solutions may be seen as directly reflecting the fact
that the su(1,1)$\sim$so(2,1) algebra associated with the
solutions \cite{lgfcav01} becomes doubled and eventually leads to
a larger so(2,2) potential algebra of the model~\cite{lgfcav02}.
In fact, the doubling of the states is one of the key motivations of 
our present analysis. It indicates that instead of one SUSY partner,
${\cal PT}$ symmetric potentials can have two, i.e. one with $q=1$
and another with $q=-1$. Similar results have been obtained for
the ${\cal PT}$ symmetrized spiked harmonic oscillator in 
\cite{mzjpa02}, however the visualization and transparency of the
example were significantly obscured there by the presence of the
essential singularity of the solutions at $x=0$ \cite{das}. 
The smooth Scarf II potential is much better suited for the similar 
purposes.

We found that the two ``fermionic'' partner Hamiltonians are 
isospectral with the ``bosonic'' one in the sense that the spectrum 
of ${\bf H}^{(q)}_+$ (${\bf H}^{(-q)}_+$) misses the level 
corresponding to $E^{(q)}_{0,-}$ ($E^{(-q)}_{0,-}$) level of the 
``bosonic'' Hamiltonian ${\bf H}_-$. This 
isospectrality has been demonstrated explicitly in the case of the 
Scarf II potential, but we also showed that it holds for any 
potential characterized simultaneously by ${\cal PT}$ symmetry and 
supersymmetry. The situation is illustrated schematically in figure 1. 
For the Scarf II potential this network of interrelated levels 
is practically the same as the one obtained in terms of an so(2,2) 
potential algebra \cite{lgfcav02}, and the two sets of SUSY shift 
opetarors correspond to the two sets of so(2,1) ladder operators. 
It has to be stressed that although the spectra of 
Hamiltonians with opposite quasi-parity are interrelated here, 
there is no operator which would flip the $q$ of a given 
wavefunction, so the two quasi-parity sectors remain disjoint in 
this sense. 

The results obtained for the Scarf II potential have significantly 
different
implications for unbroken and broken ${\cal  PT}$ symmetry. In the
former case the ``fermionic" partner potentials (\ref{vpq}) are
${\cal  PT}$ symmetric, and the $E^{(q\alpha,\beta)}_{n,+}$ energy
eigenvalues remain real. In the latter case, however, the coupling
parameters of both the even and odd component of the potential
become complex due to the imaginary value of $\alpha$, therefore
the ``fermionic" potentials cease to be ${\cal  PT}$ symmetric.
This manifest breakdown of ${\cal PT}$ symmetry is also demonstrated 
by the fact that the complex energy eigenvalues cease to appear in 
complex conjugated pairs in the spectrum of the ``fermionic'' 
Hamiltonians, because the equivalent of the 
$E^{(\pm q\alpha,\beta)}_{0,-}$ ``bosonic'' state will be missing 
from the spectrum of the ${\bf H}^{(\pm q)}_+$ ``fermionic'' Hamiltonian. 

We note that a generalized form of the ${\cal  PT}$ invariant
Scarf II potential can be obtained by an {\it imaginary} 
shift of the axis of coordinates, $x\rightarrow x+{\rm i}\epsilon$
\cite{lgmz00,shi}. Such a transformation cannot influence any of
the conclusions of our work since the modified potential and
eigenfunctions behave in the same way under the ${\cal  PT}$
transformation as the original ones, only their functional form
becomes more complicated when we decide to stay on the real line
and change the variables accordingly. This clarifies why the
energy eigenvalues are independent of $\epsilon$ as well as why
the change may only influence the wavefunctions and/or the
reflection and transmission coefficients \cite{lgfcav01}.

Turning now to more general considerations, we note that 
in the ${\cal  PT}$
symmetric setting the supersymmetrization can be realized in an
alternative way, as discussed in \cite{qtsusy}. In such a
framework the SUSY algebra is to be realized by the operators that
contain the antilinear ${\cal T}$ operation explicitly. One 
need not even use any particular potential to reveal the relation of
this scheme to the conventional one \cite{AMb}. It suffices to
recollect that the SUSY charge and shift operators may contain
the time reflection (i.e., complex conjugation) operator ${\cal 
T}$, say, in the form
\begin{equation}
\widetilde{Q}=\left(
\begin{array}{cc} 0 & 0 \\
{\cal  T} A^{(q)} & 0 \\
\end{array}\right)
\hskip 1.5cm
\widetilde{Q}^{\dag}=\left(
\begin{array}{cc} 0 & A^{\dag (q)}{\cal  T} \\
0 & 0 \\
\end{array}\right)
\label{qtscharge} .
\end{equation}
Consequently, the SUSY Hamiltonian is different in its
``fermionic" component
\begin{equation}
\widetilde{\cal  H}=\left(
\begin{array}{cc} \widetilde{H}^{(q)}_- & 0 \\
0 & \widetilde{H}^{(q)}_+ \\
\end{array}\right)
\equiv
\left(
\begin{array}{cc}  A^{\dag (q)}A^{(q)} & 0 \\
0 &  {\cal  T} A^{(q)}A^{\dag (q)} {\cal  T} \\
\end{array}\right)\ .
\label{qtsham}
\end{equation}
This indicates that the ``bosonic" component of the modified
Hamiltonian is the same as in the original case (\ref{sham}),
$\widetilde{H}^{(q)}_-=H^{(q)}_-$, while the ``fermionic'' component 
of the modified Hamiltonian coincides with the complex conjugate of 
the original ``fermionic" Hamiltonian $\widetilde{H}^{(q)}_+={\cal 
T}H^{(q)}_+{\cal  T}$. Introducing the shifted energy scale as in 
(\ref{hmeig}) and (\ref{hpeig}) these relations become 
$\widetilde{\bf H}_-={\bf H}_-$ and 
$\widetilde{\bf H}^{(q)}_+={\cal T}{\bf H}^{(q)}_+{\cal  T} 
-[\varepsilon^{(q)}]^*$. 
For unbroken ${\cal PT}$ symmetry of ${\bf H}_-$, i.e. when 
the energy eigenvalues are real and consequently 
$\varepsilon^{(q)}$ is also real, this means that the energy 
eigenvalues of $\widetilde{\bf H}^{(q)}_+$ are also real, while 
for spontaneously broken ${\cal PT}$ symmetry, when the energy 
eigenvalues and $\varepsilon^{(q)}$ are complex, the energy 
eigenvalues of $\widetilde{\bf H}^{(q)}_+$ are the complex 
conjugates of the eigenvalues of ${\bf H}^{(q)}_+$. 
The eigenfunctions are equally trivially related to the original
``fermionic" eigenfunctions in both cases.

Furthermore, the ${\cal  PT}$ invariance leads to a special
relation between the ${\cal  P}$ and ${\cal  T}$ operations
themselves. If ${\bf H}^{(q)}_+$ is ${\cal  PT}$ symmetric, then the 
complex conjugation operation has the same effect on it as the 
${\cal  P}$ spatial reflexion operation, so $\widetilde{\bf H}^{(q)}_+$ 
contains the spatially reflected potential appearing in 
${\bf H}^{(q)}_+$, so the modified SUSY construction does not differ 
essentially from the usual one. A similar relation holds between the 
eigenfunctions, if they are eigenfunctions of the ${\cal  PT}$ operator, 
i.e. if the ${\cal PT}$ symmetry is unbroken. The energy eigenvalues of
$\widetilde{\bf H}^{(q)}_+$ are real and the same as those of 
${\bf H}^{(q)}_+$, 
as we have seen above. In the case of spontaneously broken ${\cal  PT}$ 
symmetry the situation is different since the eigenfunctions are not 
invariant under the ${\cal  PT}$ operation anymore. The energy 
eigenvalues remain the same since the complex conjugate pairs simply 
transform into themselves under complex conjugation. However, in the 
case of the spontaneously broken ${\cal  PT}$ symmetry, the ${\cal  PT}$
invariance of $\widetilde{\bf H}_-={\bf H}^{(q)}_-$ need not lead to the 
${\cal PT}$ invariance of $\widetilde{\bf H}^{(q)}_+$ 
(and thus to that of ${\cal T}U^{(q)}_+(x){\cal T}=U^{(-q)}_+(x)$, 
as we have seen on the example of the Scarf II potential), so the 
whole SUSY construction can break down in this case.

\section{Summary}

In general, solvable ${\cal  PT}$ symmetric potentials may have a
richer spectrum than their Hermitian counterparts. {\it A priori},
this feature may have non-trivial implications in the SUSY
constructions. We investigated the ${\cal  PT}$ symmetric version
of the Scarf II potential in the role of a ``bosonic'' potential
and described in detail a construction of its ``fermionic" SUSY
partners.

Our first finding was based on the familiar knowledge that our
potential possesses two sets of normalizable eigenfunctions
distinguished by their quasi-parity quantum number $q=\pm 1$. On
this basis we arrived at our first important observation that one
can introduce {\em two different} superpotential functions
which lead to {\em two different} SUSY partners of the original
potential.

The formal application of the standard rules of 
supersymmetric quantum mechanics requires the vanishing
ground-state energy of the ``bosonic" potential, so a relative
energy shift of the $q=1$ and $q=-1$ sectors is needed to
correlate the energy scales. (This corresponds to switching to 
the Hamiltonians ${\bf H}_-$ and ${\bf H}^{(q)}_+$ instead of 
$H^{(q)}_-$ and $H^{(q)}_+$.) With this shift the ``bosonic"
potential can be made independent of the quasi-parity quantum
number $q$. We found that the two partner potentials are Scarf II
potentials with the parameters $\pm q\alpha+1$ and $\beta+1$. Their
energy spectrum contains one less normalizable state than the
``bosonic" potential, and the missing level carries the 
same quasi-parity quantum number as the ``fermionic'' Hamiltonian 
${\bf H}^{(q)}_+$. 

This is similar to the structures found within the standard
SUSYQM, although the related concept of the shape invariance must
be modified slightly. Still, the situation becomes perceivably
different for the unbroken and for the spontaneously broken ${\cal 
PT}$ symmetry. In the former case, both $\alpha$ and $\beta$ and
the energy eigenvalues are real. In this case both ``fermionic" 
partner potentials exhibit ${\cal PT}$ symmetry. In contrast,
for spontaneously broken ${\cal PT}$ symmetry (i.e. when $\alpha$
is imaginary, $\beta$ is real and the energy eigenvalues form
complex conjugate pairs) the ${\cal PT}$ symmetry of the
``fermionic" partner potentials may become broken {\em
manifestly}. Our second important observation is that the
isospectrality of the potentials (with the exception of the
``bosonic" $n=0$ states) still holds in this latter case, too.

Additionally, we considered an alternative SUSY construction for
${\cal PT}$ symmetric potentials in the spirit of 
\cite{qtsusy} where the SUSY algebra has been realized by SUSY
charge operators containing the antilinear operator ${\cal T}$ in
an explicit form. Our considerations were not restricted to a
particular potential, rather their validity was universal. Again,
the ``bosonic" Hamiltonian was the same as in the standard SUSYQM
approach to ${\cal  PT}$ symmetric potentials, while the
``fermionic" partner Hamiltonians (with $q=\pm 1$) proved to be
the complex conjugates of the original ``fermionic" Hamiltonians.
If the ``fermionic" potentials possessed the ${\cal  PT}$ symmetry
(as was the case of with the ``bosonic" Scarf II potential with
unbroken ${\cal  PT}$ symmetry), then the ${\cal T}$ operation had
the same effect on them as ${\cal  P}$. They simply proved to be
the spatially reflected versions of the ``fermionic" partner
potentials in the standard SUSYQM setting. Their spectrum was the
same, therefore, and the wavefunctions were related to the
original ``fermionic" wavefunctions in a trivial way. It has to be 
emphasized, nevertheless, that the isospectrality with the
original ``fermionic" potential holds even when the ``fermionic"
potentials are not ${\cal  PT}$ symmetric (as was the case for the
``bosonic" Scarf II potential with spontaneously broken ${\cal 
PT}$ symmetry).

\ack
This work was supported by the OTKA grant No. T031945 (Hungary)
and by the GA AS grant No. A 104 8004 (Czech Republic).

\newpage

\begin{figure}[h]
\setlength{\unitlength}{1cm}
\begin{picture}(  15.00,  11.00)( -2, -0.5)

%
\put(2.5,1.5){\makebox(0,0){${\bf H}^{(-q)}_+$}}%
%
%
\put(1.8,2.5){\makebox(0,0){$-q$}}
%
\put(1.5,5.2){\line(1,0){0.6}}
\put(1.5,6.2){\line(1,0){0.6}}
\put(1.5,7.2){\line(1,0){0.6}}
%
%
\put(2.8,2.5){\makebox(0,0){$q$}}
\put(2.5,3.5){\line(1,0){0.6}}
\put(2.5,4.5){\line(1,0){0.6}}
\put(2.5,5.5){\line(1,0){0.6}}
\put(2.5,6.5){\line(1,0){0.6}}
\put(2.5,7.5){\line(1,0){0.6}}

\put(4,1.5){\makebox(0,0){$\longleftarrow$}}
\put(4,1.0){\makebox(0,0){$A^{(-q)}$}}
\put(4,8.5){\makebox(0,0){$\longrightarrow$}}
\put(4,9.0){\makebox(0,0){$A^{\dag (-q)}$}}

%
%
\put(5.5,1.5){\makebox(0,0){${\bf H}_-$}}%
%
%
\put(4.8,2.5){\makebox(0,0){$-q$}}
\put(4.5,4.2){\line(1,0){0.6}}
\put(4.5,5.2){\line(1,0){0.6}}
\put(4.5,6.2){\line(1,0){0.6}}
\put(4.5,7.2){\line(1,0){0.6}}
%
%
\put(5.8,2.5){\makebox(0,0){$q$}}
\put(5.5,3.5){\line(1,0){0.6}}
\put(5.5,4.5){\line(1,0){0.6}}
\put(5.5,5.5){\line(1,0){0.6}}
\put(5.5,6.5){\line(1,0){0.6}}
\put(5.5,7.5){\line(1,0){0.6}}

\put(7,1.5){\makebox(0,0){$\longrightarrow$}}
\put(7,1.0){\makebox(0,0){$A^{(q)}$}}
\put(7,8.5){\makebox(0,0){$\longleftarrow$}}
\put(7,9.0){\makebox(0,0){$A^{\dag (q)}$}}
%
%
\put(8.5,1.5){\makebox(0,0){${\bf H}^{(q)}_+$}}%
%
%
\put(8.8,2.5){\makebox(0,0){$q$}}
\put(7.5,4.2){\line(1,0){0.6}}
\put(7.5,5.2){\line(1,0){0.6}}
\put(7.5,6.2){\line(1,0){0.6}}
\put(7.5,7.2){\line(1,0){0.6}}
%
\put(7.8,2.5){\makebox(0,0){$-q$}}
%
\put(8.5,4.5){\line(1,0){0.6}}
\put(8.5,5.5){\line(1,0){0.6}}
\put(8.5,6.5){\line(1,0){0.6}}
\put(8.5,7.5){\line(1,0){0.6}}

\end{picture}
\caption{
Schematic illustration of the relation between the spectra of 
the ``bosonic'' Hamiltonian ${\bf H}_-$ and its two ``fermionic'' 
partners ${\bf H}^{(q)}_+$ and ${\bf H}^{(-q)}_+$. In all three 
spectra levels with quasi-parity $q$ ($-q$) are degenerate with 
the corresponding levels in the other potentials, except that 
the lowest level with $q$ ($-q$) is missing from the spectrum of 
the ``fermionic'' Hamiltonian ${\bf H}^{(q)}_+$ (${\bf H}^{(-q)}_+$). 
The levels of ${\bf H}^{(q)}_+$ (${\bf H}^{(-q)}_+$) are connected 
with those of the ``bosonic'' Hamiltonian ${\bf H}_-$ by the 
$A^{(q)}$ and $A^{\dag (q)}$ ($A^{(-q)}$ and $A^{\dag (-q)}$) 
SUSY shift operators. The energy scale and the relative spacing 
of the energy levels is arbitrary.
}
\label{hosp}
\end{figure}
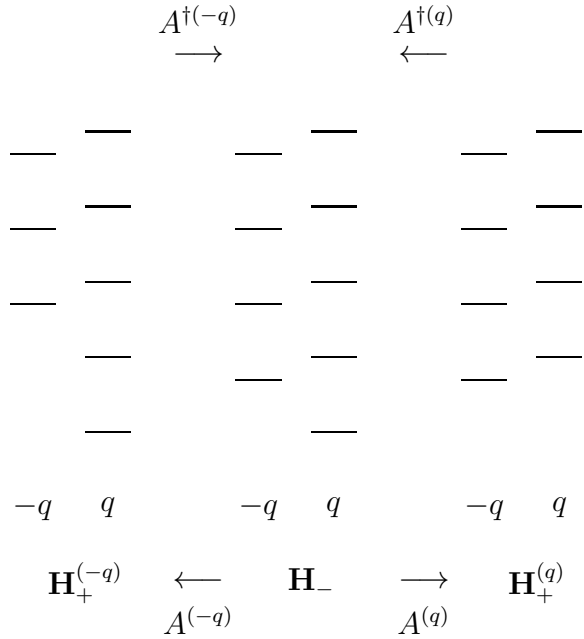

\end{document}